\documentclass{article}

\usepackage[english]{babel}

\usepackage[letterpaper,top=2cm,bottom=2cm,left=3cm,right=3cm,marginparwidth=1.75cm]{geometry}

\usepackage{amsmath}
\usepackage{graphicx}
\usepackage{algorithm}
\usepackage[noend]{algorithmic}
\algsetup{linenosize=\small}
\usepackage[colorlinks=true, allcolors=blue]{hyperref}

\title{An efficient distributed scheduling algorithm for relay-assisted mmWave backhaul networks}
\author{Qiang Hu, Yuchen Liu, Yan Yan, Miao Liu, Jun Zheng, and Douglas M. Blough}

\begin{document}
\maketitle

\begin{abstract}
In this paper, a novel distributed scheduling algorithm is proposed, which aims to efficiently schedule both the uplink and downlink backhaul traffic in the relay-assisted mmWave backhaul network with a tree topology. The handshaking of control messages, calculation of local schedules, and the determination of final valid schedule are all discussed. Simulation results show that the performance of the distributed algorithm can reach very close to the maximum traffic demand of the backhaul network, and it can also adapt to the dynamic traffic with sharp traffic demand change of small-cell BSs quickly and accurately.
\end{abstract}

\section{Introduction}
In 5G and B5G era, to achieve better cellular performance, operators are
deploying more and more small-cell base stations (BSs) in the urban area
~\cite{rong20185g,liu2018millimeter}. This brings the backhaul challenge
for the dense small-cell deployment that huge amount of cellular data 
need to be transmitted between BSs in such networks~\cite{feng2016millimetre,jaber20165g}.  
However, most of small-cell BSs may not have wired connections to the 
core network due to the construction limits and prohibitive cost in many 
places such as North America, so that mmWave backhauling has been regarded 
as a promising solution to the backhaul challenge in 5G cellular systems
since it is able to support multi-Gbps wireless backhaul links~\cite{taori2015point,shariat2015enabling,polese2020integrated}.
Currently, a lot of researches in this area focus on the mmWave integrated 
access and backhaul (IAB) network architecture where the access tier and the 
backhaul tier share the same frequency band~\cite{polese2020integrated,saha2018integrated,madapatha2020integrated}.
However, it could be challenging for the construction of line-of-sight (LOS)
mmWave backhaul links in such networks, since the location and deployment
of small-cell base stations are usually pre-fixed, which concerns more
from the perspective of increasing the cellular coverage and performance.
If the LOS path between a pair of BSs is blocked by the abundant obstacles
in the dense urban area, due to the blockage effect~\cite{liu2019analysis,gerasimenko2019capacity},
the capacity of the corresponding logical link will drop severely. 
Therefore, we proposed a new relay-assisted mmWave backhaul network
architecture~\cite{hu2017relay,hu2018optimizing,hu2020feasibility,yan2021load}, 
where dedicated mmWave relays are introduced as additional
network elements in the backhaul network. With such simple relay devices,
not only the blockage can be addressed through flexible relay deployment
in the 3D space~\cite{liu2019joint,yan2018path,liu2018blockage}, but 
also the link capacity of each physical link can be improved because of 
the shorter physical link length. 

In our previous work~\cite{hu2020feasibility},
we find that to support the ultra-high data rate in the 5G small-cell backhaul,
not only the location of relays need to be carefully selected, but also the
scheduling of the logical links between BSs have to be smartly coordinated.
Our recent work~\cite{hu2022maximum} focuses on optimizing the 
throughput performance of the relay-assisted mmWave backhaul network with 
a tree topology in either the downlink or uplink case. 
However, the scheduling algorithm proposed in~\cite{hu2022maximum} only 
aims to show the feasibility of finding a schedule that is able to satisfy 
the set of schedule lengths of logical links obtained from solving the 
optimization problems. 
It is not an algorithm that can be directly deployed into the practical 
mmWave backhaul network system. Moreover, in the performance optimization, 
the link schedule length is a continuous value, but in the 
realistic system, time resource is partitioned into small slots with the 
same duration. Furthermore, in the practical cellular network, since the 
aggregated traffic load may be larger than the network capacity, the backhaul 
traffic has to be scheduled proportionally to the traffic demand of each 
small-cell BS, so that a certain level of fairness can be achieved. Besides,
similar to the discussion in~\cite{hu2022maximum}, network
level intelligent scheduling has to come into play to handle the scenario 
where mutual interference exists between logical links and limited number of 
radio chains are available on BSs. Therefore, it is important to address the 
scheduling problem in the practical backhaul system.

In this paper, we propose a novel distributed scheduling algorithm that
can be deployed to efficiently schedule the data traffic in the practical 
relay-assisted mmWave backhaul network with a tree topology. Different 
from the work in~\cite{hu2022maximum}, where either 
downlink or uplink traffic is considered, the proposed distributed algorithm 
is able to schedule both downlink and uplink traffic in every subframe. Since
the mutual interference relationship between a pair of logical links may vary
according to different transmission directions of both links, we assume 
that two logical links are considered always interfering with each other if 
there exists at least one combination of transmission directions which leads
to significant mutual interference when both logical links transmit concurrently. 
Although applying this assumption sacrifices a certain amount of throughput 
performance, it reduces the complexity of the algorithm design. In fact, since 
there are only a few pairs of logical links interfering with each other in the 
relay-assisted mmWave backhaul network in the dense urban area when the relay 
locations are carefully selected, the throughput performance loss is likely to 
be limited.  

In the following sections, we first give an overview on the distributed 
scheduling algorithm, which describes the general idea of the algorithm design. 
Then, details on the system setting are provided. After that, three main 
components of the algorithm, i.e., the handshaking procedure for exchanging 
control information, the calculation of the local schedule of a small-cell BS, 
and the determination of the final valid schedule of a BS, are elaborated one 
after another. Simulation results show that the average throughput performance of 
the distributed scheduling algorithm can reach closely to the calculated 
maximum traffic demand of small-cell BSs (see~\cite{hu2022maximum}) 
in the tree-style mmWave backhaul network. Moreover, simulations are also
conducted to show that the distributed algorithm is able to adapt to the 
dynamic traffic demand (with sharp changes) of BSs in the network as well.

\section{Related work}
Scheduling algorithm design for the mmWave backhaul network has attracted a 
significant amount of interest in the research academia in the last decade. 
The authors of~\cite{arribas2019optimizing} study the mmWave self-backhaul 
scheduling problem and derive an MILP formulation for it as well as upper 
and lower bounds. They prove that the problem is NP-hard and can be approximated, 
but only if interference is negligible. 
Given a set of mm-wave backhaul links, ~\cite{saad2019millimeter} addresses the problem of backhaul scheduling through considering the mutual interference and number of radio chains constraints. A succinct optimization-based formulation of the problem is provided and using reduction from the set-cover problem, the authors devise a provably good polynomial-time algorithm for the problem.
The authors of~\cite{niu2019relay} investigate the problem of optimal scheduling to maximize the number of flows satisfying their QoS requirements with relays exploited to overcome blockage, where relays refer to small-cell BSs. Both a relay selection algorithm and a transmission scheduling algorithm are proposed to increase the network throughput through addressing the blockage issue and exploiting concurrent transmissions. 
~\cite{garncarek2019mmwave} studies the mmWave wireless backhaul scheduling problem in the mmWave IAB backhaul network, assuming stochastic arrival of packets at the macro-cell BS to be delivered to small-cell BSs. The authors present various results concerning system stability, defined as a bounded expected queue sizes of macro-cell BS and small-cell BSs, under different patterns of random traffic.
Note that all the works above consider the self-backhaul network where mmWave small-cell BSs serve as relays, and none of them take dedicated mmWave relays as network elements into account. In contrary, when dedicated relays are introduced in our network model, the scheduling problem becomes more sophisticated. Moreover, we also propose a new distributed scheduling algorithm for urban mmWave backhaul networks that takes several factors about practical deployment into account.

The authors in~\cite{fang2021joint} optimize the scheduling of access and backhaul links such that the minimum throughput of the access links is maximized based on the revised simplex method. ~\cite{ranjantwo} considers the radio resource scheduling problem and presents a QoS-based downlink scheduler designed explicitly for IAB networks. The scheduler is devised after considering multihop relaying topology, QoS requirements and backhaul constraints. These works are different from ours because they consider an IAB network structure whereas the frequency band used for backhaul in our research is dedicated to the backhaul usage.

In~\cite{yuan2018optimal}, the authors propose an efficient scheduling method, so-called schedule-oriented optimization, based on matching theory that optimizes QoS metrics jointly with routing in the mmWave cellular network. It is claimed to be capable of solving any scheduling problem that can be formulated as a linear program whose variables are link times and QoS metrics. As an example, they also show the optimal solution of the maximum throughput fair scheduling (MTFS).
Recently, the authors of~\cite{sha2022near} investigate two aspects of the near interference-free (NIF) space-time user scheduling in a multi-cell mmWave network with multi-RF-chain base stations. Firstly, they study the NIF user scheduling problem to minimize the unfulfilled user requirements; secondly, they study the joint NIF user scheduling and power allocation problem to minimize the total transmit power under the constraint of rate requirements.
The above works do not consider a relay-assisted backhaul network architecture, and focuses on the scheduling optimization in the mmWave access network.

~\cite{li2017practical} proposes Distributed Maximum QoS-aware (DMQ) scheduling algorithm for the mmWave backhaul network of small cells to maximize the system throughput while satisfying QoS requirements for each flow. Based on CSMA/CA, the proposed algorithm is the first that prioritizes MAC contention window to provide better concurrent transmission support while achieving QoS-aware capability. Different from our work, this paper consider a network architecture without dedicated mmWave relays deployed, and some practical factors such as the number of radio chains on BSs are not considered as well.

More recently, due to the fast development of artificial intelligence, several learning based scheduling methods for mmWave networks are also proposed. To achieve resilience to link and node failures, the authors in~\cite{dogan2021reinforcement} explore a state-of-the-art Soft Actor-Critic (SAC) deep reinforcement learning algorithm, that adapts the information flow through the network, without using knowledge of the link capacities or network topology. This work uses an N-relay Gaussian Full-Duplex (FD) 1-2-1 network model which is different from our network model. ~\cite{zhang2021resource} proposes a reinforcement learning approach based on column generation to address the resource allocation problem for mmWave IAB backhaul, as the method is claimed to be able to capture the environment dynamics such as moving obstacles in the network. ~\cite{gupta2020learning} is a work investigating the multihop link scheduling problem with the objective of minimizing the end-to-end delay experienced by a typical packet. The authors model the system as a network of queues and formulate it as a Markov decision process over a continuous action space. This allows them to leverage the deep deterministic policy gradient algorithm from reinforcement learning to learn the delay minimizing scheduling policy under two scenarios.

\section{Algorithm overview}
The basic idea of the proposed algorithm is that in the relay-assisted mmWave 
backhaul network with a tree topology, every BS except the leaf small-cell BSs, 
makes scheduling decisions for all the logical links between the BS and its child 
BSs based on several types of information, including the traffic demand information, 
the queue length information, the final valid schedule received from its parent BS, 
and the set of local schedules received from its child BSs. All information is  
collected through monitoring the BS's own status and exchanging control 
messages between the BS and its one-hop neighbors via the handshaking 
procedure introduced later. 

The link scheduling happens on a per-subframe basis, and each subframe 
contains a fixed number of time slots with the same duration. Therefore, 
both the local schedule and the final valid schedule record the assignment 
of time slots to a set of logical links for each subframe. In the distributed 
scheduling algorithm, before a BS starts to calculate its final valid schedule 
for a subframe, it has to know its parent BS's final valid schedule for the 
same subframe in advance, because the schedule of the logical link between the 
BS and its parent BS is specified therein. This indicates a strong ``happen-before" 
relationship in the calculation of final valid schedule. Since the calculation and 
propagation of the final valid schedules cannot complete instantly across 
the entire network, every BS calculates its final valid schedule for a specific
subframe at least one subframe before the that subframe starts. Thus, in 
a tree topology, the BSs closer to the macro-cell BS will always make the
scheduling decision for a future subframe earlier than those BSs farther 
away from the macro-cell BS. However, the calculation of local schedule 
does not need any information with a strict ``happen-before" relationship. 
Therefore, a BS calculates its local schedule at the beginning of a 
subframe, and send it to its parent BS immediately after the calculation 
completes, so that the received local schedule can be used by the parent 
BS to calculate its final valid schedule within the same subframe. Details 
can be found in the section~\ref{sec:ch6-handshaking}.

Moreover, in the calculation of both types of schedules, the local fairness is 
addressed, which means a BS assigns time resource to logical links 
proportionally to their traffic demand or queue lengths. This will be 
elaborated in later sections as well. 

\section{System settings}
In a given relay-assisted mmWave backhaul network, there is one 
macro-cell BS $M$, a set of small-cell BSs $\mathcal{B}=\{B_1, B_2, ...\}$, 
and a set of mmWave relays. $M$, $\mathcal{B}$, and the set of logical 
links $\mathcal{L}$ between them together form a tree topology, where 
$M$ is the root node of the tree. 

A logical link $L_i\in \mathcal{L}$ ends at the BS $B_i$, and it is either a 
single-hop path or a multi-hop path going through several mmWave relays sequentially. 
Every physical link within a logical link is a LoS. Every logical link
is ``intra-path" interference-minimal, which means the mutual interference between
the concurrent transmissions on the physical links within a logical link can be 
ignored. Based on our previous work, the optimal scheduling of a interference-minimal 
logical link can be computed.  If a logical link $L_j$ is attached to a BS $B_i$ (i.e.,
$L_j$ is between $B_i$ and $B_j$), the physical link directly attached to $B_i$ within
$L_j$ is named as $l_{ji}$. Therefore, the portion of time used by $l_{ji}$ in the 
optimal schedule length of $L_j$ is denoted as $P_{ji}$. The data rate of physical link
$l_{ji}$ is $r_{ji}$, which is a known system parameter.

It has to be clarified that the interference relationship between a pair of logical links
in the backhaul network may vary according to the combination of their transmission directions in a downlink and uplink hybrid case. If there are $x$ pairs of logical links that may interfere with each other, in total there could be $4^x$ different interference cases in the network, which introduces huge complexity to address them all. To address this issue, we assume that as long as two logical links with certain directions interfere with each other, they are considered always interfering with each other no matter which directions of their transmissions are. Under this assumption, despite that the throughput performance can not reach the optimal, we could first schedule both downlink and uplink traffic together, and then separate the traffic of different directions into different numbers of slots according to the traffic demand or queue length information.

The number of mmWave radio chains available on $B_i$ or $M$ is $N_{i}^R$ or $N_M^R$, respectively. There are 2 control time slots followed by $n_d$ data time slots in each subframe.

\section{Main components of the distributed scheduling algorithm}
In is section, we explain the distributed algorithm in details. The explanation 
focuses on three main components of the algorithm: the handshaking of control messages, 
the calculation of local schedule, and the determination of final valid 
schedule.
\subsection{Handshaking of control messages}
\label{sec:ch6-handshaking}

\begin{figure}[ht]
  \centering
  \includegraphics[width=1\linewidth]{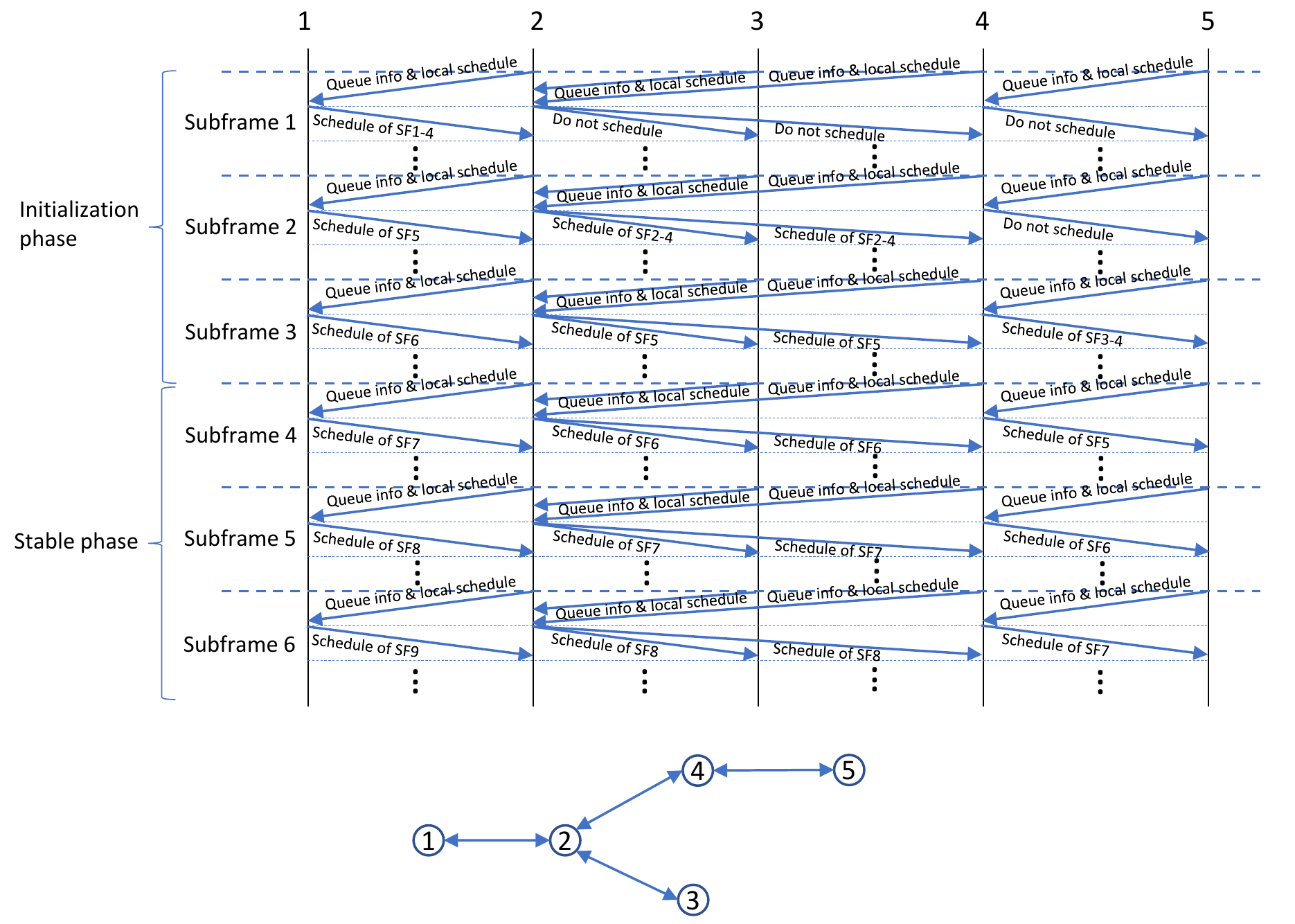}
  \caption{Hand-shaking procedure of the distributed scheduling algorithm}
  \label{fig:handshake}
\end{figure}

To elaborate the handshaking process, an example is provided in Figure~\ref{fig:handshake}. A small backhaul network with a tree topology is shown in the lower part of Figure~\ref{fig:handshake}, where BS 1 is the macro-cell BS, and BS 2-5 are all small-cell BSs. 

It is noticed that some logical links between BSs are multi-hop relay paths. If control 
messages are sent through the multi-hop logical links, the delay of these control information will be too long. To address this issue, each BS is equipped with a communication module working at a lower frequency band (e.g., sub-6 Ghz band), which uses an omni-directional antenna. That communication module takes responsibility of transmitting and receiving control messages in the first two control slots of each subframe. To avoid the collision of the control messages between a BS and its child BSs, each control slot is further partitioned into $n_{sub}$ sub-slots, and each child BS is assigned a unique sub-slot to exchange control messages. The assignment of sub-slots between a BS and its child BSs happens in the initial configuration phase when the relay-assisted backhaul network is constructed. During the initial configuration phase, a child BS is associated with its parent BS to form the backhaul logical link. 

As we can see in Figure~\ref{fig:handshake}, at the first slot of each subframe, each small-cell BS calculates its local schedule based on the traffic demand information, which will be explained in detail later. The calculated local schedule and updated queue and traffic demand information of a small-cell BS will be sent immediately to its parent BS. Note that, since there is no parent BS of the macro-cell BS, it will skip this step.

After a BS receives the control messages from its child BSs, before it starts the calculation of a final valid schedule for a specific future subframe, it checks whether the final valid schedule from its parent BS for the same future subframe has been received or not. If it has not been received, the BS will not schedule transmissions in that future subframe, because it does not know which slots are assigned to the logical link between itself and its parent BS. Any transmission from the BS may potentially conflict with the transmission from its parent BS. On the other hand, if the interested final valid schedule from the parent BS has been received, the BS will starts the calculation of its final valid schedule.
  
As shown in the example, each BS will start its scheduling in a specific subframe according to its height in the tree topology. If the topology has a depth of $H$, which is the height of the macro-cell BS and each small-cell BS $B_i$'s height is denoted as $h_i$, the macro-cell BS will start its scheduling in subframe 1, while small-cell BS $B_i$ will start its scheduling in subframe ($H-h_i+1$). Note that, because the schedule of leaf BSs is determined by their parent BSs, they do not calculate their final valid schedule. 

Moreover, in the first scheduling subframe, each non-leaf BS will calculate its schedule of the next several subframes in advance (i.e., from subframe $H-h_i+1$ to subframe $H$), because when a child BS calculates its final valid schedule of a subframe, it has to know its parent BS's final valid schedule of the same subframe in advance; however, the schedule needs time to propagate to its child BSs. After the first scheduling subframe, each BS only needs to calculate the final valid schedule of the next unscheduled subframe. As depicted in Figure~\ref{fig:handshake}, 
the initialization phase lasts $H-1$ subframe, and after that, the \textit{stable phase} starts.


\subsection{Calculate the local schedule of small-cell BSs}
\label{sec:local_schedule}
The purpose of calculating the local schedule on a small-cell BS $B_i$ is to obtain the ``desired" number of slots that $B_i$ wants its parent BS $B_{i}^p$ to assign for the data transmission on the logical link $L_i$ between $B_i$ and $B_i^p$. As a ``desired" local optimal, it is calculated based on the local traffic demand information rather than the queue information. 

Traffic demand information is an upper layer (i.e., application) statistics which is not 
typically used in the scheduling algorithm (i.e., MAC layer) in wireless ad hoc or mesh networks. However, in cellular networks starting from the 4G-LTE systems, the Quality-of-Service (QoS) has become a key performance metric.
QoS related policies such as traffic classification and prioritization have already been installed in BSs. Application or service data rate of each user accessing to a BS, as a crucial QoS parameter, is constantly collected by the BS. In the 5G system, as a wider range of applications are going to be supported, even finer-grained management of QoS is expected to be implemented, which means more accurate traffic demand information is likely available at each BS~\cite{ferrus20185g}. In fact, the local schedule aims to provide a ``desired" resource allocation requirement to the parent BS, the amount of which is usually much larger than the amount of resource that 
the parent BS can offer in a backhaul network with the heavy traffic load. From this point of view, a small estimation error on the traffic demand of each BS is acceptable. From the discussion in the above section~\ref{sec:ch6-handshaking}, we know that each BS keeps updating its traffic demand information to its parent BS; thus, after a certain amount of time, BS $B_i$ can obtain the accurate total traffic demand $D_j$ from its child BS $B_j \in \mathcal{B}_{i}^c$, where $\mathcal{B}_i^c$ is the set of child BSs of $B_i$. Note that, first, the traffic demand $D_j$ of $B_j$ is in fact the 
aggregated traffic demand of all small-cell BSs in the sub-tree rooted at $B_j$; second, as the proposed distributed algorithm schedules both uplink and downlink traffic, the traffic demand information contains both the downlink traffic demand and the uplink traffic demand information.

The local schedule of $B_i$ indicates the number of slots assigned for the data transmissions (both uplink and downlink) on all physical links attached to $B_i$. We denote the number of slots assigned to the attached physical link within the logical link $L_j$ as $n_j$. 
Note that, the demand may exceed the maximum throughput capacity of a BS; therefore, we use a local continuous scale variable $S_i$ to indicate the fraction of traffic demand of each child BS actually being served at $B_i$ in one subframe. From this perspective, our proposed distributed scheduling algorithm addresses the local fairness through proportionally schedule the data traffic according to the amount of the traffic demand of each child BS. 

Based on the above system setting, we can formulate a mixed integer programming problem to 
address the calculation of the local scheduling on $B_i$, in which the local scale variable $S_i$ 
is to be maximized. In Equation~\ref{eq:local_schedule}, continuous variable $S_i$ is the 
optimization objective, and integer variables $\{n_j|\ B_j \in B_i \bigcup \mathcal{B}_{i}^c\}$ 
are auxiliary variables.
\begin{equation}
\begin{split}
    \max \quad & S_i \\ 
    \mathrm{s.t.} \quad & n_i\cdot r_{ii} \geq S_i \cdot \sum_{B_j \in B_i \bigcup \mathcal{B}_{i}^c}{D_j} \\
    & 0 \leq S_i \leq 1 \\
    & n_j \cdot r_{ji} \geq S_i\cdot D_j,\ \forall\ B_j \in \mathcal{B}_{i}^c\\
    & \frac{1}{P_{ji}} \leq \frac{n_j}{P_{ji}} \leq n_d,\ \mathrm{if}\ D_j > 0, \ \forall\ B_j \in B_i \bigcup \mathcal{B}_{i}^c \\
    & {n_j} = 0, \ \mathrm{if}\ D_j = 0, \ \forall\ B_j \in B_i \bigcup \mathcal{B}_{i}^c \\
    & \lceil\frac{n_j}{P_{ji}}\rceil + I_{jk}\cdot{\lceil\frac{n_k}{P_{ki}}\rceil} \leq n_d,\ \forall\ B_j, B_k \in B_i \bigcup \mathcal{B}_{i}^c \\
    & \sum_{B_j \in B_i \bigcup \mathcal{B}_{i}^c}{n_j} \leq n_d\cdot N_i^R 
\end{split}
\label{eq:local_schedule}
\end{equation}

As the distributed scheduling algorithm is expected to operate in the real relay-assisted mmWave backhaul system, in which every physical link has a length in the order of 200 m. Due to the use
of high-gain highly-directional antennas, the data transmissions on these LoS physical links are 
likely to be able to use the highest level of modulation (e.g., 256 QAM) allowed in the system.
Therefore, the physical link data rate is the same, and we denote the unified physical link data 
rate as $r$. Since the portion $P_{ji}$ of the transmission on the physical link within the logical link $L_j$ attached to $B_i$ is either 1 (i.e., single-hop) or 0.5 (i.e., multi-hop), we define a new integer parameter $\alpha_{ji}= \frac{1}{P_{ji}}$, whose value is either 1 or 2. Therefore, we can update the above mixed integer programming problem into a mixed integer linear programming problem. 
\begin{equation}
\begin{split}
    \max \quad & S_i \\ 
    \mathrm{s.t.} \quad & n_i\cdot r \geq S_i \cdot \sum_{B_j \in B_i \bigcup \mathcal{B}_{i}^c}{D_j} \\
    & 0 \leq S_i \leq 1 \\
    & n_j \cdot r \geq S_i\cdot D_j,\ \forall\ B_j \in \mathcal{B}_{i}^c\\
    & \alpha_{ji} \leq \alpha_{ji}\cdot {n_j} \leq n_d, \ \mathrm{if}\ D_j > 0, \ \forall\ B_j \in B_i \bigcup \mathcal{B}_{i}^c \\
    & {n_j} = 0, \ \mathrm{if}\ D_j = 0, \ \forall\ B_j \in B_i \bigcup \mathcal{B}_{i}^c \\
    & \alpha_{ji}\cdot {n_j} + I_{jk}\cdot{\alpha_{ki}\cdot {n_k}} \leq n_d,\ \forall\ B_j, B_k \in B_i \bigcup \mathcal{B}_{i}^c \\
    & \sum_{B_j \in B_i \bigcup \mathcal{B}_{i}^c}{n_j} \leq n_d\cdot N_i^R 
\end{split}
\label{eq:local_schedule_MILP}
\end{equation}

Since in the real system, $\mathcal{B}_i^c$ is a small set, $n_d$ is a pre-set small integer, the known interference relationship matrix $\{I_{jk}\}$ is sparse, the computation time for solving this mixed integer linear programming problem is very short. After solving the problem, the maximum $S_i$ is found, and correspondingly, we can calculate the minimum number of time slots $\widehat{n_i}$ to support the maximum achievable traffic demand for the physical link attached to $B_i$ within $L_i$. $\widehat{n_i}$ is the key parameter of the local schedule, which will be transmitted to $B_i$'s parent BS, as it indicates the maximum time slots that the parent BS shall allocate to the physical link within $L_i$. The parent BS could allocate more time slots to $L_i$; however, the extra traffic cannot be absorbed timely by $B_i$, as the total traffic on $L_i$ exceeds the maximum achievable traffic demand of $B_i$.

In the real deployment of the distributed algorithm, since the traffic demand of each small-cell BS
may be fluctuating within a small range in most of time or it changes gradually in a slow speed, the calculation of local schedule may fall into the trouble of tracking the demand change too sensitively, such that the final valid schedule may be changing too frequently as well. To address this issue, we modify the message exchange of local schedule that the BS $B_i$ only updates the value of $\widehat{n_i}$ to its parent BS when significant changes of $\widehat{n_i}$ happens. Specifically, $B_i$ keeps recording the 
temporary $\widehat{n_i}$ value calculated in each subframe, and it applies a sliding window on the sequence of $\widehat{n_i}$ values to get the time-averaged value $\bar{\widehat{n_i}}$. When the difference between the current reporting $\widehat{n_i}$ value and the time-averaged value $\bar{\widehat{n_i}}$ is larger than a threshold of $T$ percentage, the reporting $\widehat{n_i}$ value is updated to $\bar{\widehat{n_i}}$; otherwise the reporting value does not change. In our simulations, we choose a relatively large threshold (i.e., 50\%), so that in a small backhaul network (e.g., 20 BSs), few small-cell BSs will experience sharp traffic demand change simultaneously.

\subsection{Determine the final valid schedule of a BS}
After a BS receives the queue, traffic and local schedule information from its child BSs, it will determine its schedule of a 
future subframe based on its height in the topology, as mentioned in the above section on hand-shaking. 
Different from the calculation of local schedule, the valid schedule is determined using not only the traffic demand information, but also the queue and local schedule information. 

To realize the distributed algorithm, a non-leaf small-cell BS $B_i$ has to maintain queues 
to store downlink and uplink packets, whose destinations are not $B_i$. We use $\mathcal{B}_i$
to denote the set of BSs located in the sub-tree rooted at $B_i$.
For each small-cell BS $B_k\in\mathcal{B}_i$, $B_i$ maintains a downlink queue and a 
uplink queue for it, which temporarily store all packets with the same destination BS as $B_k$ and all 
packets to be routed to the macro-cell BS $M$ with the same source BS as $B_k$, respectively. 
At the beginning of a subframe on $B_i$, the total number of packets to $B_k$ in the corresponding 
downlink queue is $q_{i,k}^D$; while the total number of packets from $B_k$ in the corresponding 
uplink queue is $q_{i,k}^U$.

However, the set of queue lengths actually used in the final valid schedule calculation are 
related to the routing, because we need to know the exact total number of uplink and downlink
packets that wait to be transmitted on each logical link $L_j$ between $B_i$ and $B_i$'s each 
child BS $B_j$. Therefore, at BS $B_i$, the total number of downlink packets waiting to be
transmitted on $L_j$ is denoted as $Q_{i,j}^D$. Meanwhile, the uplink packets waiting to be 
transmitted on $L_j$ are stored in the uplink queues at $B_j$, and the total number of them
is denoted as $Q_{j}^U$. 
\begin{equation}
    \begin{split}
        Q_{i,j}^D = \sum_{B_k\in \mathcal{B}_j}{q_{i,k}^D}\\
        Q_{j}^U = \sum_{B_k\in \mathcal{B}_j}{q_{j,k}^U}\\
    \end{split}
    \label{eq:queue_length}
\end{equation}

Based on Equation~\ref{eq:queue_length}, the total number of uplink and downlink packet waiting to be
transmitted on $L_j$ is $Q_j$, and 
\begin{equation}
    Q_j = Q_{i,j}^D + Q_{j}^U
\end{equation}
As mentioned in the ``handshaking" procedure, at the first time slot of each subframe, the child BS $B_j$
transmits the uplink queue information $Q_{j}^U$ to its parent BS $B_i$, so that $Q_j$ can be updated timely before the calculation of final schedule begins. Note that the queue maintenance on the macro-cell
BS $M$ and the leaf small-cell BS is similar to that on the small-cell BS, except that $M$ does not have uplink queues and leaf small-cell BSs do not have downlink queues. There is no parent BS of the macro-cell BS and no child BS of the leaf small-cell BSs. The process of generating the valid schedule 
of a BS varies according to the type of that BS.

\subsubsection{Macro-cell BS}
To determine the final schedule at the macro-cell BS $M$, it has to first 
find out the maximum number of packets it can transmit to each child BS
within a single subframe in the way local fairness is considered. Similar
to the procedure of calculating the local schedule, we first have to optimize
the local scaling variable $S_M$,
\begin{equation}
\begin{split}
    \max \quad & S_M \\ 
    \mathrm{s.t.} \quad & n_j \cdot r \geq S_M\cdot \min{\{Q_j,D_j\}},\ \forall\ B_j \in \mathcal{B}_{M}^c\\
    & 0 \leq S_M \leq 1 \\
    & 1 \leq  {n_j} \leq \widehat{n_j}, \ \mathrm{if}\ Q_j > 0, \ \forall\ B_j \in \mathcal{B}_{M}^c \\
    & n_j = 0, \ \mathrm{if}\ Q_j = 0, \ \forall\ B_j \in \mathcal{B}_M^c \\
    & \alpha_{j}\cdot {n_j} + I_{jk}\cdot{\alpha_{k}\cdot {n_k}} \leq n_d,\ \forall\ B_j, B_k \in \mathcal{B}_{M}^c \\
    & \sum_{B_j \in \mathcal{B}_{M}^c}{n_j} \leq n_d\cdot N_M^R 
\end{split}
\label{eq:mbs_schedule}
\end{equation}

In Equation~\ref{eq:mbs_schedule}, the first constraint guarantees that 
a logical link $L_j$ attached to the macro-cell BS will be allocated a 
certain number of slots so that $S_M\cdot Q_j$ packets 
in the queue of $L_j$ can be transmitted in the scheduled subframe, 
where $Q_j$ is the total number of packets in both uplink and downlink 
queues of $L_j$, while $\mathcal{B}_{M}^c$ is the set of child small-cell
BSs of the macro-cell BS $M$.  
The third constraint indicates that for a logical link with non-empty 
queue, at least 1 time slot will be assigned to it, and the macro-cell 
BS will not assign more than $\widehat{n_j}$ to $L_j$, as this value 
is from the local schedule of $B_j$, which is considered as the maximum
number of slots scheduled for $L_j$ expected by $B_j$.
The fourth constraint is the ``logical link interference constraint"
and the last constraint is the ``radio chain constraint", both explained
in~\cite{hu2022maximum}.

\subsubsection{Non-leaf small-cell BS}
As for a non-leaf small-cell BS $B_i$, its valid schedule of a subframe is calculated upon the received valid schedule of its parent BS, the received local schedule and uplink queue information of its child BSs, and its own downlink queue information. Similar to the case of macro-cell BS, the non-leaf small-cell BS also has to first maximize the local variable $S_i$ as well, 
\begin{equation}
\begin{split}
    \max \quad & S_i \\ 
    \mathrm{s.t.} \quad & n_j \cdot r \geq S_i\cdot \min{\{Q_j,D_j\}},\ \forall\ B_j \in \mathcal{B}_{i}^c\\
    & 0 \leq S_i \leq 1\\
    & 1 \leq {n_j} \leq \widehat{n_j}, \ \mathrm{if}\ Q_j > 0, \ \forall\ B_j \in \mathcal{B}_{i}^c \\
    & n_j = 0, \ \mathrm{if}\ Q_j = 0, \ \forall\ B_j \in \mathcal{B}_i^c \\
    & I_{ik}\cdot{\alpha_{k}\cdot {n_k}} \leq n_d - \alpha_{i}\cdot {\widetilde{n_i}},\ \forall\ B_k \in \mathcal{B}_{i}^c \\
    & \alpha_{j}\cdot {n_j} + I_{jk}\cdot{\alpha_{k}\cdot {n_k}} \leq n_d,\ \forall\ B_j, B_k \in \mathcal{B}_{i}^c \\
    & \sum_{B_j \in \mathcal{B}_{i}^c}{n_j} \leq n_d\cdot N_M^R - \widetilde{n_i}
\end{split}
\label{eq:sbs_schedule}
\end{equation}
where $\widetilde{n_i}$ is the number of slots scheduled for the 
physical link within $L_i$ attached to $B_i$, which is a known 
information for $B_i$, because it is determined and sent to $B_i$ 
by its parent BS before the calculation occurs. All the constraints
are similar to the ones in the macro-cell BS case above, except that
the total number of available time slots should subtract the ones 
(i.e., $\widetilde{n_i}$) already assigned to $L_i$.

\subsubsection{Leaf small-cell BS}
As leaf BSs do not have child BSs, their valid schedule is determined by their parent non-leaf BSs.  

After obtaining the maximum $S_i$ or $S_M$, the number of slots assigned to a logical link $L_j$ 
attached to a BS $B_i$ or $M$ can be calculated. Based on the ratio between $Q_{i,j}^D$ and $Q_{j}^U$, 
we can determine the number of slots for downlink and uplink traffic transmissions on $L_j$, respectively.
A similar scheduling procedure as the Algorithm in~\cite{hu2022maximum} can be used to actually schedule the set of slots assigned to each logical link within each subframe. 

Moreover, we have to figure out how to fill packets into each slot when the schedule is executed. Based on the slot allocation, $B_i$ (i.e., the parent of $B_j$) can get the maximum numbers of downlink packets $N_{j}^{P,D}$ and $B_j$ can get the maximum number of uplink packets $N_{j}^{P,U}$ allowed to be transmitted on $L_i$ within a subframe. Then, $B_i$ selects the downlink packets from different queues proportionally according to $\{q_{i,k}^D\}$, and $B_j$ selects the uplink packets from different queues proportionally according to $\{q_{j,k}^U\}$.

\section{Numerical results and analysis}
\subsection{The throughput performance of distributed scheduling algorithm}
In this section, simulations are conducted to evaluate the throughput performance of the proposed distributed scheduling algorithm. In the simulations, the physical link data rate is set to 13.3 Gbps as discussed in section~\ref{sec:local_schedule}. There are in total 24 slots within a subframe which lasts for 0.1 ms, and the first two slots of each subframe is reserved to implement the operation of handshaking procedure described in section~\ref{sec:ch6-handshaking}. The rest 22 slots are data slots which can be used to transmit both uplink and downlink data traffic. All backhaul topologies used in the simulations are generated using a spanning-tree algorithm
introduced in the section 4.3 of~\cite{hu2019design}.

\begin{figure}[ht]
  \centering
  \includegraphics[width=0.8\linewidth]{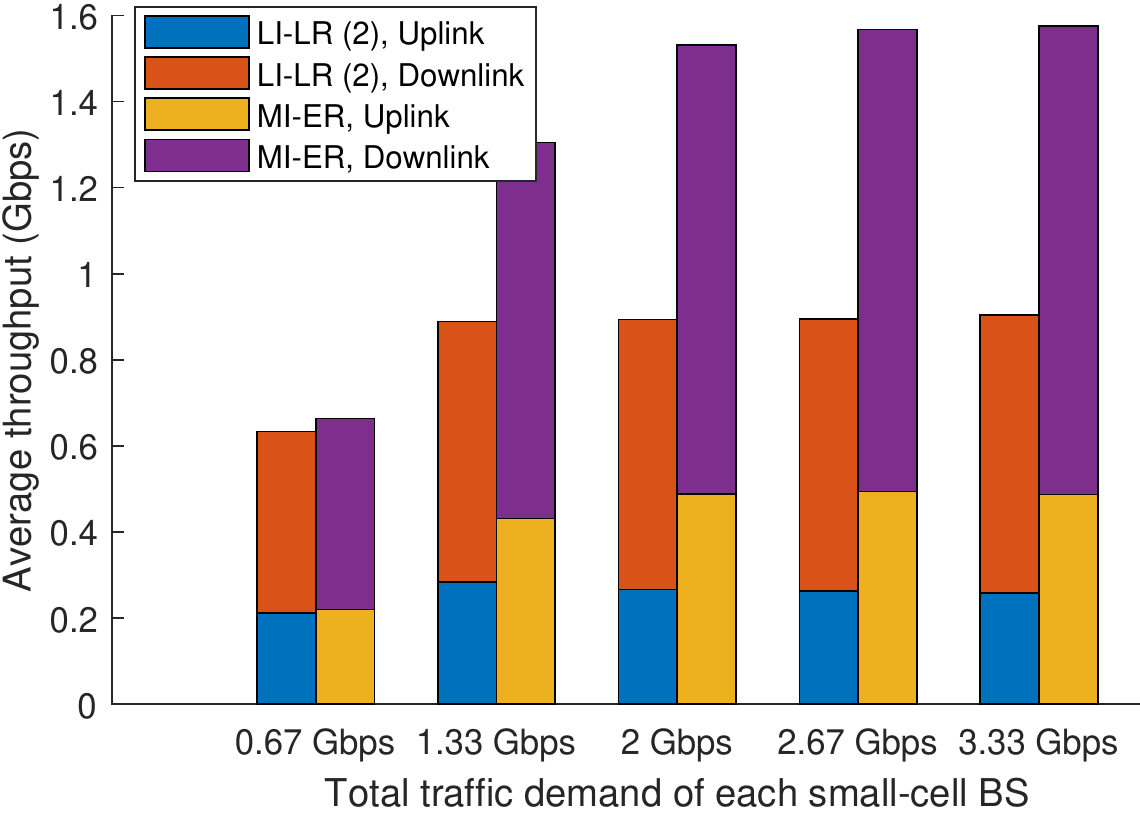}
  \caption{Throughput performance of distributed scheduling algorithm}
  \label{fig:throughput_distributed}
\end{figure}

Figure~\ref{fig:throughput_distributed} shows the throughput performance of the distributed algorithm in different network settings (i.e., the interference condition and the number of radio chains available on each BS) under different traffic loads from 0.67 Gbps to 3.33 Gbps at each small-cell BS. Note that, all the simulated throughput values are averaged across 50 sets of individual simulations in each scenario. In each simulation, the ratio between the uplink and downlink input traffic demand of a small-cell BS is 1:2, and each small-cell BS has the same traffic demand. Each individual simulation runs for 1000 subframes. In Figure~\ref{fig:throughput_distributed}, we can see that the throughput performance of the ``MI-ER" scenario (i.e., interference-minimal and enough radio chains available) is much better than that of the ``LI-LR (2)" scenario (i.e., interference exists between a few pairs of logical links, and 2 radio chains on the macro-cell BS while 1 radio chain on each small-cell BS). This is because the maximum traffic demand achievable of each small-cell BS in the ``MI-ER" case is much higher than that of the ``LI-LR (2)" case. In fact, using the updated link capacity and $\{p_i^f\}$ values, we can get the updated maximum traffic demand of each small-cell BS in the ``MI-ER" case as 1.64 Gbps in average; while the value of the ``LI-LR (2)" case is 0.94 Gbps. As Figure~\ref{fig:throughput_distributed} shows, the maximum throughput values in the simulation are close to the calculated maximum traffic demand values in both scenarios, which means the distributed algorithm can schedule the transmissions efficiently in the mmWave backhaul networks.

\subsection{Enhance the aggregated throughput achieved using the distributed algorithm}
It is noticed that the determination of the final valid schedule on BSs strictly follows the idea of 
scheduling the backhaul traffic proportionally according to the queue length of each flow bounded by
the traffic demand of each corresponding small-cell BS. Although it addresses the fairness issue, the
utilization of the network resource could be low when there exist several bottleneck routes in the 
backhaul network. On those bottleneck routes, if the aggregated traffic demand is much larger than the 
network capacity of the routes, it leads to very small values of the optimized scaling variable $S$. 
As the local $S_M$ value is applied to all routes in the network at the macro-cell BS, a very small
$S_M$ will limit the amount of traffic transferred on those non-bottleneck routes, where plenty unused
network resource may exist.

To improve the low network resource utilization and increase the network aggregated throughput, we
add a step in the determination of the final valid schedule after the optimization of $S_M$ at the 
macro-cell BS. In this new step, the optimized $\widehat{S_M}$ is used to bound smallest amount of
traffic demand of each small-cell BS that has to be serve in a subframe. In Equation~\ref{eq:mbs_schedule_modified}, we can see that the new optimization objective is to 
maximize the number of data slots used in one subframe.
\begin{equation}
\begin{split}
    \max \quad & \sum_{B_j\in \mathcal{B}_M^c}{n_j} \\ 
    \mathrm{s.t.} \quad & n_j \cdot r \geq \widehat{S_M}\cdot \min{\{Q_j,D_j\}},\ \forall\ B_j \in \mathcal{B}_{M}^c\\
    & 1 \leq  {n_j} \leq \widehat{n_j}, \ \mathrm{if}\ Q_j > 0, \ \forall\ B_j \in \mathcal{B}_{M}^c \\
    & n_j = 0, \ \mathrm{if}\ Q_j = 0, \ \forall\ B_j \in \mathcal{B}_M^c \\
    & \alpha_{j}\cdot {n_j} + I_{jk}\cdot{\alpha_{k}\cdot {n_k}} \leq n_d,\ \forall\ B_j, B_k \in \mathcal{B}_{M}^c \\
    & \sum_{B_j \in \mathcal{B}_{M}^c}{n_j} \leq n_d\cdot N_M^R 
\end{split}
\label{eq:mbs_schedule_modified}
\end{equation}

\begin{figure}[h!]
  \centering
  \includegraphics[width=0.75\linewidth]{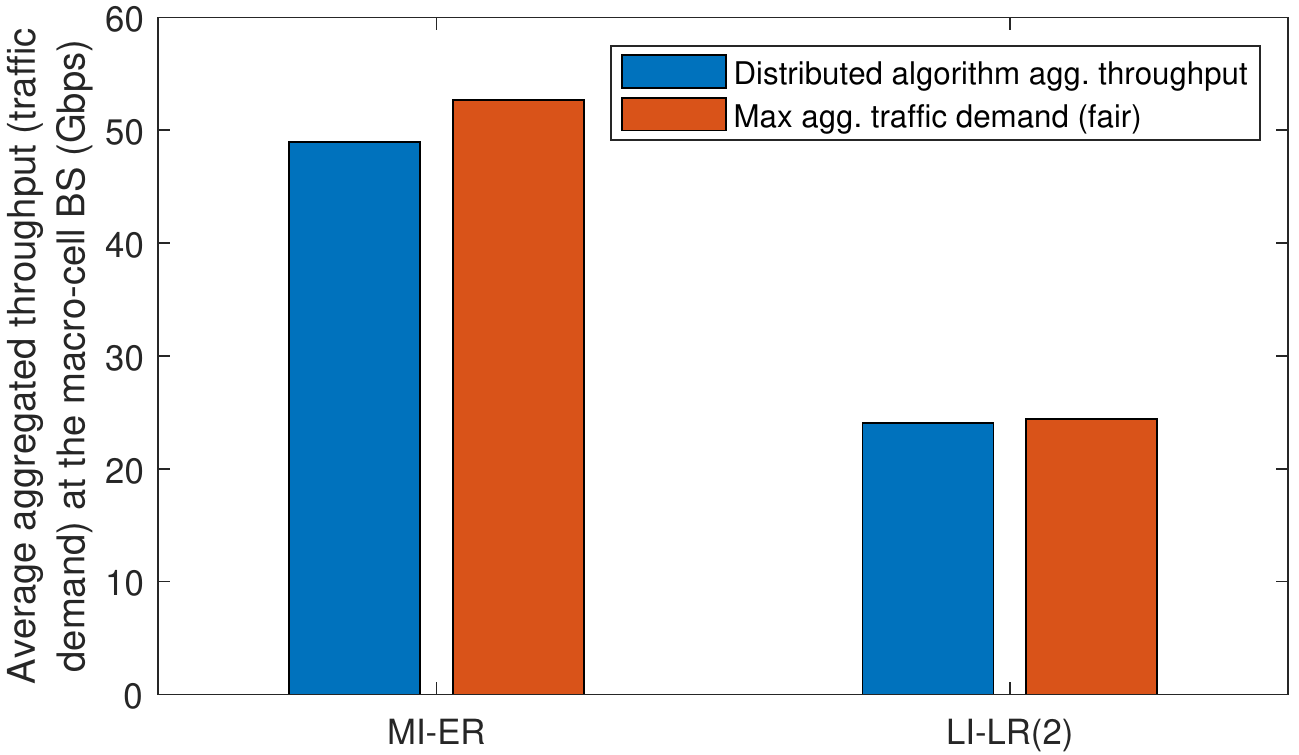}
  \caption{Comparison between the aggregated throughput achieved using the distributed algorithm and the maximized aggregated traffic demand}
  \label{fig:aggTD_dist_fair}
\end{figure}

We conduct simulations to compare the aggregated throughput achieved by using the modified distributed scheduling algorithm and the maximized aggregated traffic demand at the macro-cell BS that can be supported by the backhaul network, which is obtained in~\cite{hu2022maximum}. In our distributed scheduling simulations, the traffic demand of each small-cell BS is set as 3.33 Gbps, and the aggregated traffic demand surpasses the network capacity. 
As we can see in Figure~\ref{fig:aggTD_dist_fair}, in both MIER and LILR(2) scenarios, the aggregated 
throughput achieved using the modified algorithm is close to the maximized traffic demand that can be supported by the backhaul network. It shows that our proposed algorithm can schedule the backhaul traffic efficiently. 

\begin{figure}[h!]
  \centering
  \includegraphics[width=0.75\linewidth]{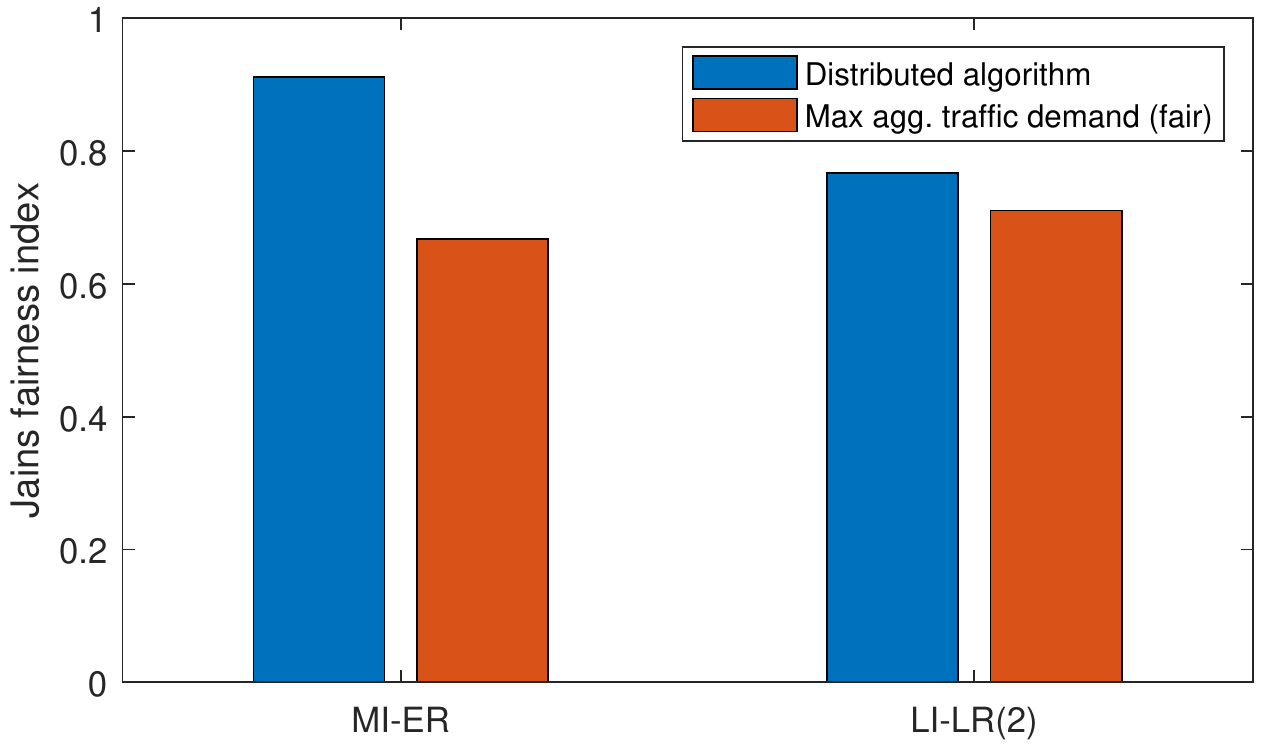}
  \caption{Comparison on the achieved fairness index values between applying the distributed algorithm and maximizing the aggregated traffic demand}
  \label{fig:fairIndex_dist_fair}
\end{figure}

Moreover, as the input traffic demand is the same on each small-cell BS, we also calculate the Jain's fairness index of the set of throughput values of all small-cell BSs obtained in the distributed scheduling simulation. Figure~\ref{fig:fairIndex_dist_fair} shows that our algorithm can provide better fairness than the case in the theoretical analysis in~\cite{hu2022maximum} can. Because in~\cite{hu2022maximum}, the optimization does not address the fairness issue in the allocation of the extra to be supported traffic demand among small-cell BSs. In the solution to the optimization problem obtained in the theoretical analysis, usually the extra traffic demand is allocated to only a few BSs, which leads to an unfair situation. In contrary, using our distributed algorithm, the extra traffic demand is allocated locally fairly due to the procedure of determining final valid schedule at small-cell BSs.

\subsection{Track the dynamic traffic demand of a small-cell BS}

\begin{figure}[h!]
  \centering
  \includegraphics[width=0.6\linewidth]{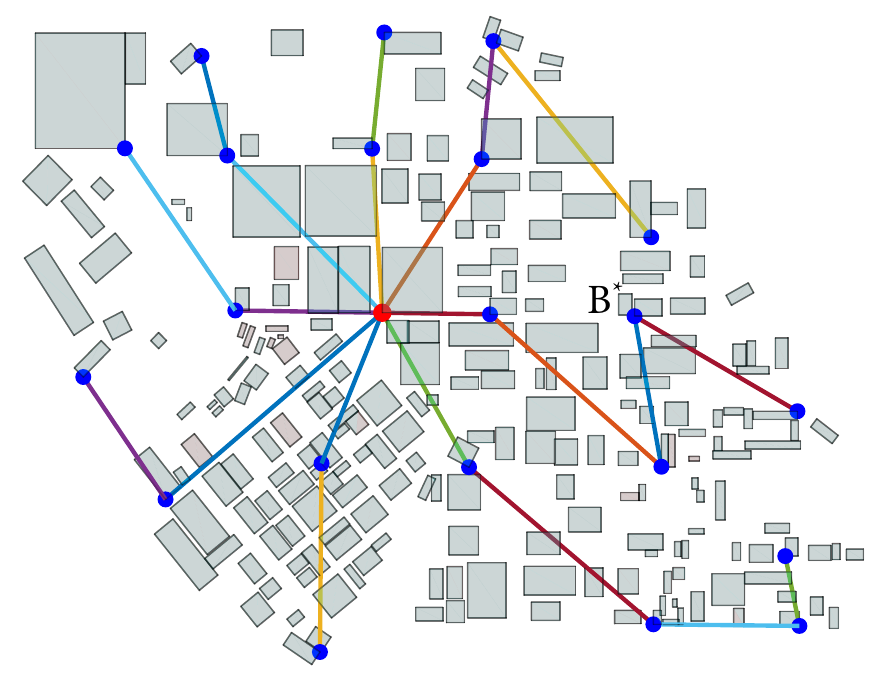}
  \caption{Backhaul topology (logical links) used in the simulation}
  \label{fig:backhaul_dynamic_traffic}
\end{figure}

In the practical backhaul system, the traffic demand of each small-cell BS may be fluctuating 
within a small range in most of the time or it changes gradually with a slow speed; therefore,
during a relatively long period of time, only a few BSs are likely to experience sharp traffic
demand change. In this simulation, we aim to explore the feature of our proposed distributed 
scheduling algorithm on tracking the dynamic traffic demand (with sharp demand change) of a 
small-cell BS in the backhaul network. The topology used in the simulation is shown in 
Figure~\ref{fig:backhaul_dynamic_traffic}, where the targeting BS is marked as $B^{*}$. 

In the simulation, the traffic demand of all small-cell BSs is initialized as 0.67 Gbps for 
downlink and 0.33 Gbps for uplink. When time arrives at the 250-th subframe, the downlink 
traffic demand of $B^*$ is doubled to 1.34 Gbps; while its uplink traffic demand is unchanged 
until the 400-th subframe. Starting from the 400-th subframe, the uplink traffic demand of 
$B^*$ is doubled. The downlink and uplink traffic demands of $B^*$ are set back to their 
initial values at the 600-th and 750-th subframe, respectively.
\begin{figure}[h!]
  \centering
  \includegraphics[width=0.8\linewidth]{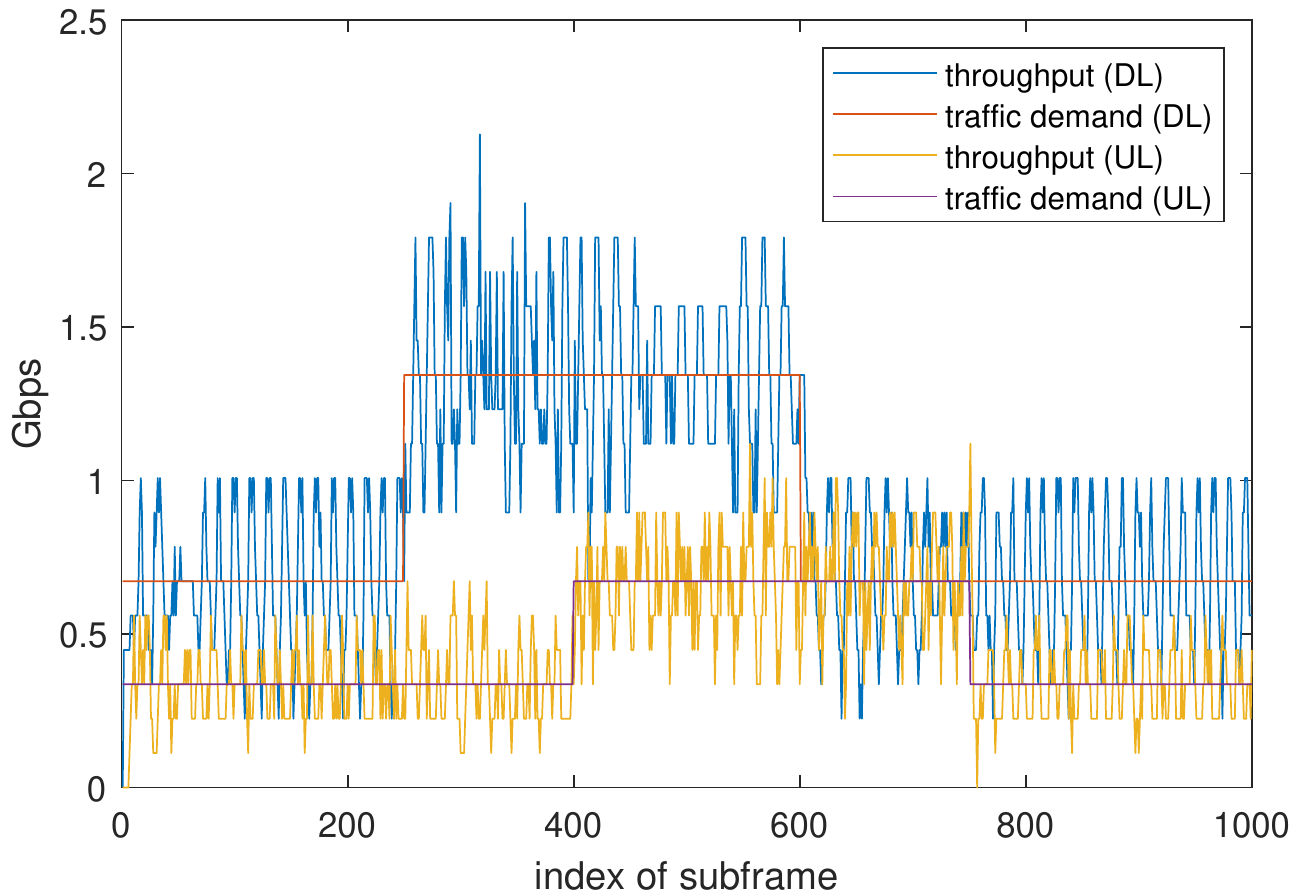}
  \caption{Dynamic throughput within each subframe at a BS}
  \label{fig:dynamic_traffic}
\end{figure}
Figure~\ref{fig:dynamic_traffic} shows the dynamic throughput within each subframe of $B^*$.  
From the figure, we can see that the proposed distributed scheduling algorithm can track the 
sharp traffic demand change quickly and accurately. In fact, when the downlink traffic 
demand doubles, it takes 8 subframes for the instantaneous downlink throughput to jump to the 
new stage; while it takes 13 subframes for the instantaneous downlink throughput to drop back 
to the old stage, after the downlink traffic demand sets back to the original value. 
Correspondingly, the two values of number of subframes used to track the uplink traffic demand 
changes are 4 and 8. The selected BS has a height of 4 in the topology. In the downlink traffic 
demand change case, the traffic demand change has to first propagate to the macro-cell BS to 
increase the ``desired" time slots for the logical links in the route. After that, more packets 
can be transferred to the targeting BS. However, in the uplink case, the increased uplink 
traffic flows together with the ``traffic increasing" message to the macro-cell BS, which is 
a one-way trip intuitively. When the traffic demand drops, it takes a bit longer time for the throughput to drop, because the packets from the heavy traffic queued in the intermediate BSs need time to be absorbed by $B^*$.

\section{Conclusions}

In this paper, a novel distributed scheduling algorithm is created, which aims to efficiently schedule both the uplink and downlink backhaul traffic in the mmWave backhaul network with a tree topology. The handshaking of control messages, calculation of local schedules, and the determination of final valid schedule are all discussed. Simulation results show that the performance of the distributed algorithm can reach very close to the aforementioned maximum traffic demand  of the backhaul network, and it can also adapt to the dynamic traffic with sharp traffic demand change of small-cell BSs quickly and accurately.

\bibliographystyle{ieeetr}
\bibliography{sample}

\end{document}